\documentclass[12pt]{article}

\usepackage{scicite}

\usepackage{times}

\usepackage{amsmath}
\usepackage{amsfonts}
\usepackage{amssymb}
\usepackage{graphicx}

\graphicspath{{pict/}{}}

\newenvironment{sciabstract}{%
\begin{quote} \bf}
{\end{quote}}

\title{Spontaneous photon-pair generation at the nanoscale}

\author
{Giuseppe Marino${}^{1,2,3\ast}$, Alexander S. Solntsev${}^{1,4}$, Lei Xu${}^{1,5,9}$, Valerio F. Gili${}^{3}$, \\ Luca Carletti${}^{6}$, Alexander N. Poddubny${}^{1,7,8}$, Mohsen Rahmani${}^{1}$, Daria A. Smirnova${}^{1}$,\\ Haitao Chen${}^{1}$, Aristide Lema\^{i}tre${}^{10}$, Guoquan Zhang${}^{9}$, Anatoly V. Zayats${}^{2}$, Costantino\\ De Angelis${}^{6}$, Giuseppe Leo${}^{3}$, Andrey A. Sukhorukov${}^{1}$, Dragomir N. Neshev${}^{1\ast}$\\
\\
\normalsize{${}^{1}$Nonlinear Physics Centre, Research School of Physics and Engineering},\\
\normalsize{Australian National University, Canberra ACT 2601, Australia}\\
\normalsize{${}^{2}$Department of Physics and London Centre for Nanotechnology,}\\
\normalsize{King's College London, Strand, London WC2R 2LS, UK}\\
\normalsize{${}^{3}$Mat\'{e}riaux et Ph\'{e}nom\`{e}nes Quantiques,}\\ 
\normalsize{Universit\'{e} Paris Diderot-CNRS, F-75013 Paris, France}\\
\normalsize{${}^{4}$School of Mathematical and Physical Sciences,}\\
\normalsize{ University of Technology Sydney, 15 Broadway, Ultimo NSW 2007, Australia}\\
\normalsize{${}^{5}$School of Engineering and Information Technology,} \\
\normalsize{University of New South Wales, Canberra ACT 2600, Australia}\\
\normalsize{${}^{6}$Dept. of Information Engineering,}\\
\normalsize{University of Brescia, Via Branze 38, 25123 Brescia, Italy}\\
\normalsize{${}^{7}$ITMO University, 49 Kronverksky Pr., Saint Petersburg 197101, Russia}\\
\normalsize{${}^{8}$Ioffe Physical-Technical Institute, Saint Petersburg 194021, Russia}\\
\normalsize{${}^{9}$The MOE Key Laboratory of Weak Light Nonlinear Photonics,}\\
\normalsize{School of Physics, Nankai University, Tianjin, China}\\
\normalsize{${}^{10}$Centre de Nanosciences et de Nanotechnologies,}\\
\normalsize{Universit\'{e} Paris-Saclay-CNRS, Palaiseau, France}\\
\normalsize{$^\ast$ E-mail: Giuseppe.Marino@univ-paris-diderot.fr, Dragomir.Neshev@anu.edu.au}
}

\date{\today}

\begin{document} 


\baselineskip24pt

\maketitle 
\newpage

\begin{sciabstract}
Optical nanoantennas have shown a great capacity for efficient extraction of photons from the near to the far-field, enabling directional emission from nanoscale single-photon sources. However, their potential for the generation and extraction of multi-photon quantum states remains unexplored. Here we demonstrate experimentally the nanoscale generation of two-photon quantum states at telecommunication wavelengths based on spontaneous parametric down-conversion in an optical nanoantenna. The antenna is a crystalline AlGaAs nanocylinder, possessing Mie-type resonances at both the pump and the bi-photon wavelengths and when excited by a pump beam generates photon-pairs with a rate of 35~Hz. Normalized to the pump energy stored by the nanoantenna, this rate corresponds to 1.4~GHz/Wm, being one order of magnitude higher than conventional on-chip or bulk photon-pair sources. Our experiments open the way for multiplexing several antennas for coherent generation of multi-photon quantum states with complex spatial-mode entanglement and applications in free-space quantum communications and sensing.
\end{sciabstract}

\newpage


\section*{Introduction}
Correlated photon-pairs are essential building blocks for photon entanglement ~\cite{Muller:2014-224:NPHOT,Versteegh:2014-5298:NCOM}, which underpins many quantum applications, including secure networks, enhanced measurement and lithography, and quantum information processing ~\cite{OBrien:2009-687:NPHOT}.
One of the most versatile techniques for the generation of correlated photons is the process of spontaneous parametric down-conversion (SPDC)~\cite{Kwiat:1995-4337:PRL}. 
The latter SPDC allows for an arbitrary choice of energy and momentum correlations between the generated photons, robust operation at room temperature, as well as for spatial and temporal coherence between simultaneously pumped multiple SPDC sources. 

Alternative approaches based on atom-like single photon emitters, such as solid-state fluorescent atomic defects~\cite{Sipahigil:2016-847:SCI}, quantum dots~\cite{Senellart:2017-1026:NNANO, Somaschi:2016-340:NPHOT}, and 2D host materials~\cite{Aharonovich:2016-631:NPHOT, Tran:2016-37:NNANO}, have reached a high degree of frequency indistinguishability, purity and brightness~\cite{Somaschi:2016-340:NPHOT, Senellart:2017-1026:NNANO}. However, this comes with the expense of operation at cryogenic temperatures and lack of spatial coherence between multiple quantum emitters. These features might limit possible applications and reduce the potential for device scalability. Furthermore, the small size of the atomic sources often requires complex schemes aimed at coupling to optical nanoantennas and improving the photon extraction efficiency~\cite{Aharonovich:2016-631:NPHOT}. 


The miniaturization of SPDC quantum-light sources to micro and nanoscale dimensions is a continuing quest, as it enables denser integration of functional quantum devices. Traditionally, bulky cm-sized crystals were utilized for SPDC, entailing the difficulty of aligning multiple optical elements after the SPDC crystal, while offering relatively low photon-pair rates~\cite{Kwiat:1995-4337:PRL}. As a first step of miniaturization, SPDC was realized in low-index-contrast waveguides, which allowed confining light down to several square micrometers transversely to the propagation direction, significantly enhancing the conversion efficiency~\cite{Fiorentino:07}. However, this approach still requires centimetres of propagation length, which makes the on-chip integration with other elements challenging~\cite{Solntsev:2017-19:RPH}. The introduction of high-index contrast waveguides and ring resonators allowed for shrinking the sizes necessary for SPDC to millimetres~\cite{Orieux:2013-160502:PRL}, and to tens of micrometers~\cite{Guo:2017-e16249:LSA}. However, further miniaturisation down to the nanoscale requires conceptually different approaches.

For a long time, plasmonic nanoantennas have been considered as a favorable platform for enhancing single-photon emission~\cite{Schuller:2010-193:NMAT,Curto:2010-930:SCI} and nonlinear interactions~\cite{Kauranen:2012-737:NPHOT, Celebrano:2015-412:NNANO, Baselli:2017-1595:PLS, Marino:2018-1700189:LPR, Lee:2014-65:NAT}. However, the limited volume of the plasmonic modes, the losses and the centrosymmetric nature of plasmonic materials, result in a relatively low second order nonlinear conversion efficiency.
Dielectric nanoantennas have thus emerged as a an alternative nanoscale nonlinear platform~\cite{Shcherbakov:2014-6488:NANL, Yang:2015-7388:NANL, Grinblat:2017-953:ACSN, Carletti:2015-26544:OE}. The strong enhancement of the nonlinear processes observed in them is largely due to the absence of material absorption and the excitation of Mie-type bulk resonances ~\cite{Kuznetsov:2016-846:SCI}. The highest conversion efficiency to date has been achieved employing III-V semiconductor nanostructures, such as AlGaAs which is a non-centrosymmetric material with high quadratic nonlinear susceptibility. In particular, second-harmonic generation efficiencies up to $10^{-4}$ have been recently demonstrated~\cite{Gili:2016-15965:OE, Liu:2016-5426:NANL, Camacho-Morales:2016-7191:NANL, Carletti:2017-114005:NANT, Liu:2018-2507:NCOM}, six orders of magnitude higher than in plasmonics.

Despite the rapid recent progress, the generation of quantum light with nonlinear nanoantennas has not been reported to date. Such nanoscale multi-photon quantum sources would offer an unexplored avenue for applications of highly indistinguishable and spatially reconfigurable quantum states, through the spatial multiplexing of several coupled nanoantennas. Until now, the big question on whether a single nanoscale antenna can generate measurable pairs of photons with non-classical polarization and energy correlations remains open.

Here, we demonstrate experimentally the generation of spontaneous photon pairs from a single AlGaAs disk nanoantenna exhibiting Mie-type resonances at both pump and bi-photon wavelengths. As such, the generation of photon pairs is a result of the correlations between these two sub-wavelength magnetic dipole modes. The observed photon-pair generation rate is 35 Hz, which, per unit volume, is higher than other SPDC light sources. Our SPDC source offers room temperature operation and the possibility of both engineering the radiation pattern and obtaining coherent interference between multiplexed sources, thanks to its inherent capacity to shape the subwavelength electromagnetic mode fields.


\section*{Results and Discussions}

\begin{figure*}
\begin{center}
\includegraphics[width=0.9\columnwidth] {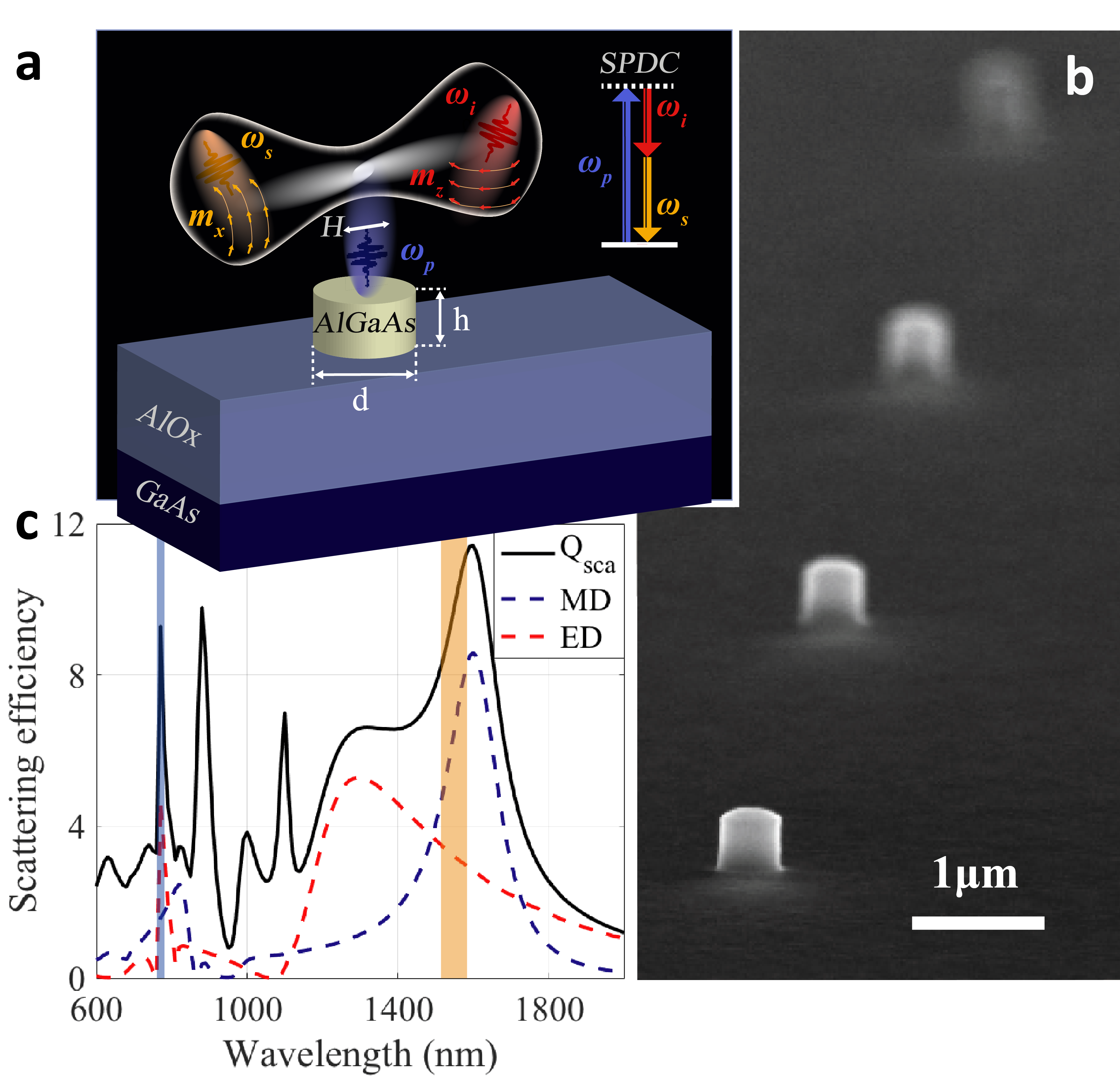}
\caption{{\bf Nonlinear nanoantenna for generation of heralded photons.}
(a) Schematic representation of the nanoantenna-based source of photon-pairs through the SPDC process. The inset depicts the energy diagram of the SPDC process. The SPDC pump is horizontally polarized along the $[100]$ AlGaAs crystallographic axis. The signal and idler photons are generated by entangled magnetic dipole moments inside the volume of the AlGaAs nanoantenna, namely $m_x$ and $m_z$ (sketched below the emitted photons). (b) A typical scanning electron micrograph (SEM) images of $[100]$ AlGaAs monolithic nanocylinders, $10\,\mu$m apart, such that each disk can be excited individually. (c) Simulated scattering efficiency, $Q_{sca}$, and multipolar decomposition in terms of the two leading electric (ED) and magnetic dipoles (MD) for a nanocylinder with diameter $d=430$~nm and height $h=400$~nm. The vertical blue and orange bars show the spectral ranges of the pump light and the generated SPDC light (signal and idler), respectively.}
\label{fig_concept}
\end{center}
\end{figure*}

A schematic of our nanoantenna photon-pair source is shown in Fig.~\ref{fig_concept}a. The nanoantenna is a crystalline AlGaAs cylinder with diameter $d=430$~nm and height $h=400$~nm. The scanning electron microscope (SEM) image of the fabricated structure is shown in Fig.~\ref{fig_concept}b. The non-centrosymmetric crystalline structure of AlGaAs offers strong bulk quadratic susceptibility of $d_{14}=100$~pm/V. The AlGaAs also exhibits high transparency in a broad spectral window from 730~nm up to the far infra-red due to the direct electronic bandgap, further preventing two-photon absorption at telecommunication wavelengths.

In our experiments, a pump beam at a frequency $\omega_p$ is focused on the nanocylinder, resulting in the simultaneous emission of signal and idler photons via SPDC. These can have different frequencies ($\omega_s$ and $\omega_i$) and polarizations, can propagate at different angles with respect to the incident pump photons, and are collected in reflection geometry.
The dimensions of the nanocylinder are chosen such that it exhibits Mie-type resonances at the pump and signal/idler wavelengths. The simulated linear scattering efficiency is defined as the scattering cross section $C_{sca}$ normalized by the cross area of the nanocylinder $\pi r^2$: $Q_{sca} = C_{sca}/\pi r^2$. It is shown in Fig.~\ref{fig_concept}c along with the two leading multipolar contributions of the scattering. In the infrared region of the spectrum, where the signal and idler photon pairs are generated, the nanocylinder exhibits a magnetic dipolar resonance, which is the lowest order Mie-mode, featuring a Q-factor of nine (Fig.~\ref{fig_concept}c). For the spectral region of the pump $760-790$~nm, we have another strong resonance with a Q factor of $52$, represented by a peak in the scattering efficiency spectrum (Fig.~\ref{fig_concept}c). This is dominated by the electric dipole moment of the antenna, although it also contains higher-order multipolar contributions (not shown). The strong internal fields at the Mie-type resonances allow for strong enhancement of the nonlinear frequency mixing processes and also imposes a spectral selection for the frequencies of the generated photons.


The SPDC process in the nanocylinder can result in the emission of photon pairs with nontrivial correlations, associated with different angular and polarization components. In order to experimentally determine the optimal conditions for photon-pair generation and ultimately for optimum SPDC efficiency, usually one uses the technique of quantum state tomography~\cite{James:2001-52312:PRA}. However, due to weakness of nonlinear process, the bi-photon rate tends to be low, thereby resulting in long time acquisition of the photon counting statistics, as well as lack of correlation precision. Therefore, optimizing the experimental parameters directly through SPDC measurements is impractical and we need an alternative solution.

To solve this issue, we resort to the quantum-classical correspondence between SPDC and its reversed process, namely sum-frequency generation (SFG), where the generated sum-frequency and pump waves propagate in opposite directions to the SPDC pump, signal and idler ~\cite{Poddubny:2016-123901:PRL, Lenzini:2018-17143:LSA}. Such quantum-classical correspondence is applicable to any quadratic nonlinear structures and allows the classical estimation of the SPDC generation bi-photon rates through the relation
\begin{equation} \label{eq:1}
   \frac{1}{\Phi_{p}} \frac{d N_{\rm pair}}{dt} =
2\pi\Xi^{\rm SFG} \frac{\lambda_p^4}{\lambda_s^{3}\lambda_i^{3}}
                  \frac{c \Delta\lambda}{\lambda_s^{2}} .
\end{equation}
Here, $\Phi_p$ is the SPDC pump flux, $\lambda_p$, $\lambda_s$ and $\lambda_i$ are the pump, signal and idler wavelengths, and $\Delta\lambda$ is the nonlinear resonance bandwidth at the signal/idler wavelengths. The efficiency $\Xi^{\rm SFG}$ is given by the ratio of sum-frequency photon power to the product of incident energy fluxes at signal and idler frequencies. Detailed derivation of Eq.~(\ref{eq:1}) along with its angular- and polarization-resolved versions are given in Sec.~1.1 of Supplementary Information. We establish that the number of photon pairs generated through SPDC, in a given optical mode of the nanostructure, is proportional to the SFG amplitude of the classical signal and idler waves, propagating in the opposite direction. In this framework, we can first optimize the SFG efficiency and thus predict the bi-photon generation rates, prior to SPDC detection. Importantly, the SFG process can also be characterized for different polarizations, further optimizing the parameters for the subsequent SPDC measurements. 

\begin{figure*}[ht!]
\begin{center}
\includegraphics[width=\textwidth]{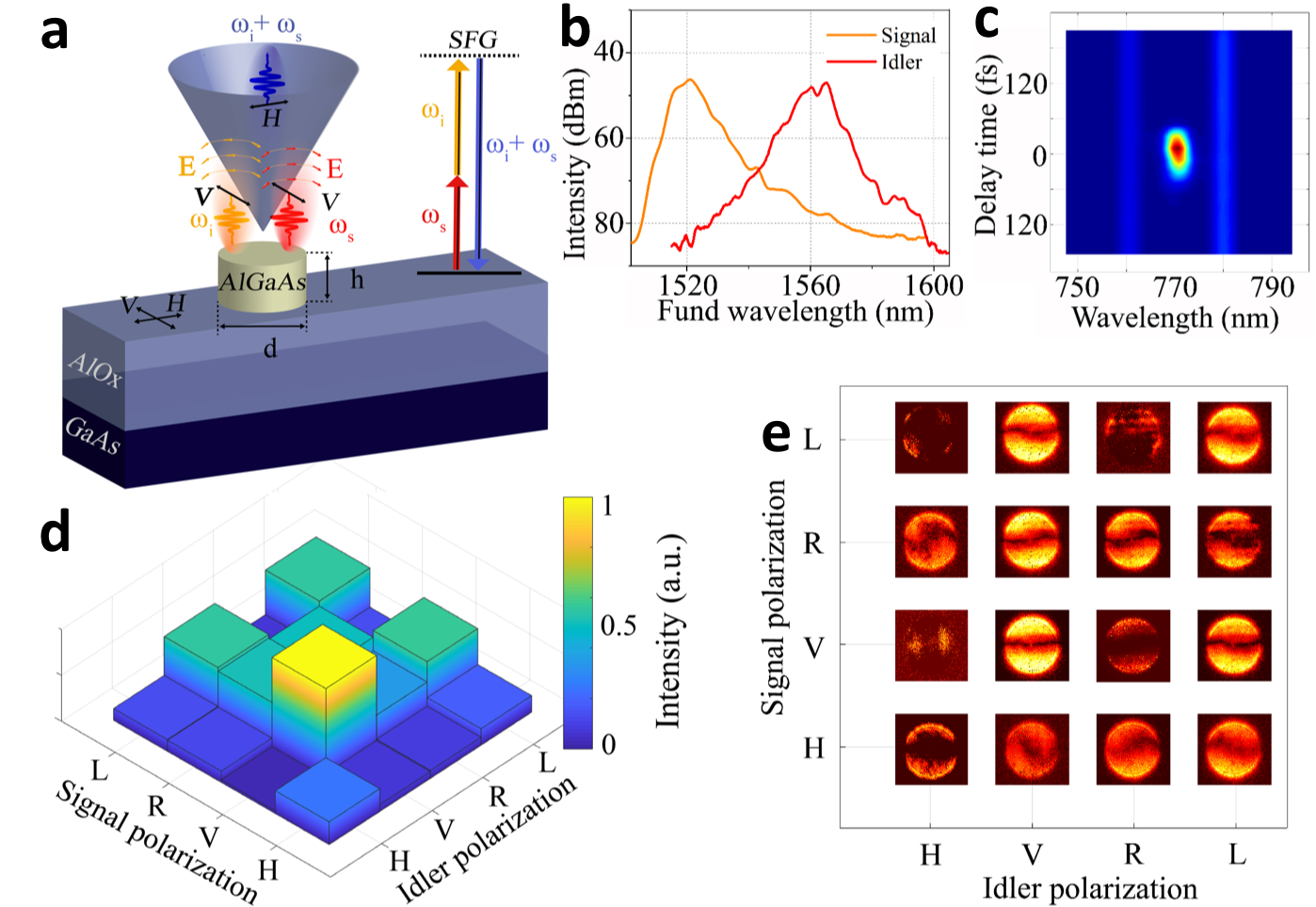}
\caption{{\bf SFG nonlinear characterisation of polarization correlations in the nanoantenna.} (a) Schematic of the experimental arrangement and energy conservation diagram of the SFG process in the inset. (b) Signal (orange line) and idler (red line) spectra filtered from the fs laser source. (c) Spectrum of the nonlinear wave mixing in the AlGaAs nanocylinder as a function of the time delay between the signal and idler pulses. The SFG only happens when the two pulses overlap, while the spectral features at 760 and 780~nm correspond to SHG from the individual signal and idler pulses. (d) Intensity of H-polarized reflected SFG at 770~nm measured with 16 combinations of horizontal (H), vertical  (V), right circular (R) and left circular (L) polarizations of signal and idler beams for the nanocylinder geometry in Fig.~\ref{fig_concept}. (e) Measured reflected SFG images in k-space for the polarization combinations shown in (d) and SFG detected with $NA=0.7$. }
\label{fig3}
\end{center}
\end{figure*}

The schematic of our SFG experiments is illustrated in Fig.~\ref{fig3}a. Two short pulses at wavelengths 1520~nm and 1560~nm illuminates the nanoantenna as signal and idler beams. Their spectra are shown in Fig.~\ref{fig3}b. The two beams are focused onto a single AlGaAs nanocylinder by a 0.7~NA objective, with 10 mW average powers, 2 $\mu$m (diameter) diffraction-limited spots and 7~GW/cm$^2$ peak intensities. The polarization in each arm can be controlled independently by half- and quarter-wave plates. The linear H polarization is parallel to the AlGaAs nano-resonator crystallographic axis $[100]$. Figure~\ref{fig3}c shows the H-polarized SFG signal collected in backward direction at different optical delays between the two V-polarized pulses. This polarizations arrangement corresponds to the maximum SFG efficiency, as we discuss below. The two spectral peaks at 760 and 780~nm correspond to the second harmonic generation (SHG) from the individual signal and idler pulses, and are observed at all delay times. The third peak at 770~nm only occurs at ``time zero,'' when the signal and idler pulses arrive at the nanocylinder simultaneously. This SFG pulse has a full-width at half-maximum (FWHM) of 80~fs, in agreement with the duration of the pump pulses.

By setting different combinations of incident polarizations for the signal and idler pulses, including horizontal (H), vertical (V), right circular (R) and left circular (L), we measured the SFG for H (or V) polarization. The choice for H polarized SFG is arbitrary, as for normally incident signal/idler beams V and H are identical due to the cylindrical symmetry of the disk and anisotropy of the material. The resulting SFG signal intensities (normalized to the maximum value, see Methods) at 770~nm and the corresponding radiation patterns recorded via a back focal plane (BFP) imaging system are shown in Figs.~\ref{fig3}d,e, respectively. The maximum signal of H-polarized SFG is obtained when both signal and idler are V-polarized. At the microscopic scale, this corresponds to the excitation of signal, idler and SFG modes whose vectorial components constructively overlap following the symmetry of AlGaAs second-order susceptibility tensor. The highest measured SFG conversion efficiency from our nanocylinder is $1.8\times10^{-5}$, which is comparable to the SHG efficiency obtained in earlier measurements~\cite{Gili:2016-15965:OE,Liu:2016-5426:NANL,Camacho-Morales:2016-7191:NANL}. 
As shown in the BFP images, the SFG radiation patterns strongly depend on signal and idler polarization combinations, however the general observation is that the SFG signal is emitted under angle, off-axis to the nanocylinder. This is due to the symmetry of the nonlinear tensor, as previously reported for SHG in Refs.~\cite{Carletti:2016-1500:ACSP, Camacho-Morales:2016-7191:NANL}. 
The experimental results were also compared with finite element simulations under realistic experimental conditions and are shown in Fig.~S3 of the Supplementary Information. The simulated SFG intensity is enhanced when the polarizations of both signal and idler beams are VV, or RR or LL polarized. Lower counts are seen for the mixed polarization cases, and for the case HH. This trend matches the experimental results, particularly for the combinations involving H and V polarizations, while the RR and LL cases appears less bright than VV case. This discrepancy can be attributed to slight non-uniformity of the fabricated structure, which has a small amount of ellipticity.

Importantly, knowing the SFG efficiency of $1.8\times10^{-5}$ for the VV $\rightarrow$ H process, we can estimate the possible bi-photon rates for detection of heralded photons from our nanoantenna SPDC source. Using Eq.~(\ref{eq:1}) we predict a photon-pair generation rate of about 380~Hz at a pump power of 2 mW. This value is significant and well above the dark count rates for our detector of 5~Hz (for details on the count rate estimation see Sec.~1.2 of the Supplementary Information). It is worth noting that in contrast to the assumption of Eq.~(\ref{eq:1}), in the experiments (see schematics in Fig.~\ref{fig_concept}a and \ref{fig3}a) all possible sum-frequency signal/idler/generation do not exactly correspond to SPDC signal/idler/pump propagating in opposite directions. Thus, some deviation of the experimentally detected rate from the value predicted above is expected. 

Detection of coincidences between photons generated through SPDC in the nanocylinder is illustrated in Fig.~\ref{fig_concept}a. 
In the experiment we use a CW pump laser, with a power of 2~mW at the wavelength of 785~nm. 
The generated photon pairs are expected to have a large spectral bandwidth of about 150~nm, due to the broad magnetic dipole resonance in the IR spectral range, as shown in Fig.~\ref{fig_concept}c. This bandwidth is quite broad with respect to conventional SPDC sources, which have typical sub-nm or few-nm bandwidth. This broad bandwidth offers a range of advantages, including a short temporal width for timing-critical measurements, such as for temporal entanglement~\cite{Rogers:2016-1754:ACSP}, or for SPDC spectroscopy~\cite{Solntsev:2018-21301:APLP}. It also dictates a sub-100 fs temporal width of the generated photons, which is much shorter than the coincidence window $\tau_c$ (see Methods).

\begin{figure}[tb]
\begin{center}
\includegraphics[width=0.99\columnwidth]{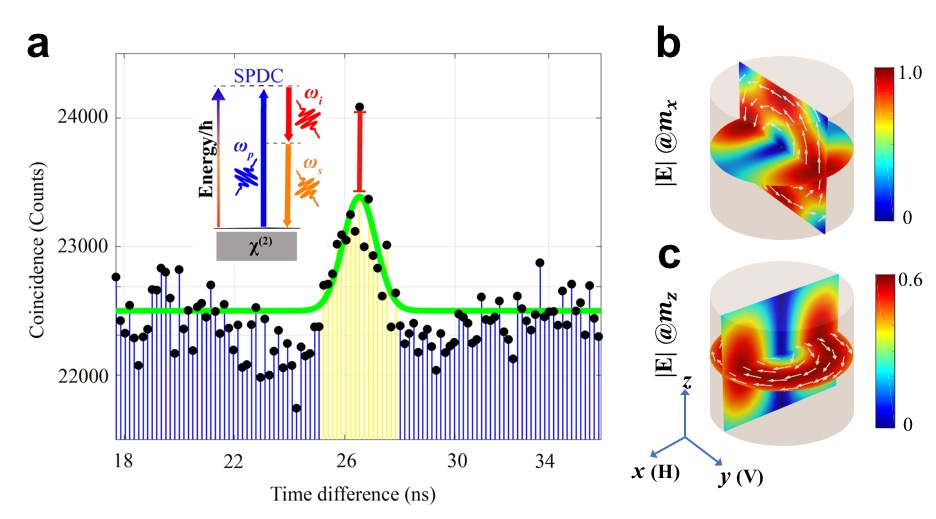}
\caption{{\bf Generation of photon-pairs in an AlGaAs disk nanoantenna.}
IR coincidence counts integrated over 24~hours on two single-photon detectors after a beam splitter (shown in Fig.~S4 of the Supplementary Information). A significant statistical increase, marked by the red bar, is apparent at a time difference of 26.5~ns, corresponding to the temporal delay between both detectors. Black dots are the measured coincidences, the yellow shadowed area indicates correlation due to thermal excitation of the semiconductor materials, while the green line is its fitted Gaussian curve. The inset shows a schematic of the SPDC process and energy correlation. (b,c) Numerically simulated fields inside the nanocylinder when exciting $m_x$ and $m_z$ modes, respectively. The white arrows indicate the electric field vector.
}
\label{fig4}
\end{center}
\end{figure}

The measured coincidences for an H-polarized CW pump are presented in Figure~\ref{fig4}a, where photon counting statistics is accumulated by integrating over 24~hours. For a time difference of 26.5~ns, corresponding exactly to the temporal delay between both detectors, we observe a single bin with high coincidence rate. This is consistent with the physics of SPDC generation of signal and idler photons with the estimated temporal correlations of sub-100~fs. Although it only emerges from the background by a limited number of counts (see Sec.~5 of the Supplementary Information), this peak of coincidence rate is statistically relevant. We also observe the indication of correlation due to thermal excitation of the semiconductor materials~\cite{Flagg:2012-163601:PRL} with approximately 2~ns width, as shown with the yellow-shadowed area and its Gaussian fit (green line) in Fig.~\ref{fig4}a. 
 
The estimation of the SPDC rate from the AlGaAs nanocylinder that takes into account the losses in the detection system (see Sec.~5 of the Supplementary Information) results in a total photon-pair generation rate from our nanoantenna of $d N^{gen}_{\rm disk} / dt = 35$~Hz. Normalized to the pump energy stored by the nanoantenna, this rate reaches values of up to 1.4~GHz/Wm, being one order of magnitude higher than conventional on-chip or bulk photon-pair sources~\cite{Kwiat:1995-4337:PRL,Guo:2017-e16249:LSA} (for details see Sec.~8 of Supplementary Information).
Importantly, this rate is significantly higher than the reference measurements of the AlOx/GaAs substrate without the nanocylinder (see Fig.~S7 of the Supplementary Information), being $dN^{gen}_{\rm sub}/dt =9$~Hz. We have also considered the likely influence of the AlGaAs nanocylinder on the SPDC from the substrate, such as refocusing the pump beam or the photon pairs, however these were found to be negligible (see Sec.~7 of the Supplementary Information).
 

Finally, we have numerically calculated the photon-correlation, shown in Fig.~S10 of the Supplementary Information, associated to the measured correlation of Fig.~\ref{fig4}a. Only orthogonal Cartesian components of magnetic dipole contribute to SPDC process, which leads to coupling of two magnetic dipole moments of the nanoantenna, namely $m_x$ and $m_z$. The two coupled modes for an H-polarised pump beam are shown in Fig.~\ref{fig4}b and c. The coupling of these two sub-wavelength modes efficiently generates photon pairs in the far-field via the antenna radiation and underpins the measured photon correlation.

\section*{Perspectives}
Our calculated correlation suggests that the photons spontaneously emitted from the AlGaAs nanoantenna might constitute pairs of polarization-entangled twin photons, i.e. they are efficiently generated when energy is conserved and spatial overlap of the corresponding  subwavelength field modes occurs.  However increase of the photon-pair rates would be required to unambiguously test the entanglement in our system.

We note that a possible increase in the rates of photon pair-generation and detection can be further anticipated by tuning the wavelength and the angle of incidence of the SPDC pump, as well as by optimising the nanoantenna geometry. Indeed, as shown in Figs.~S4,S5 of the Supplementary Information and following the discussed SFG-SPDC analogy, for a normal incident pump, the generated SPDC photons will be mostly emitted under oblique angle with respect to the nanocylinder axis. Therefore, exciting the disk with a doughnut-like cylindrical vector beam or inclined beam might result in stronger generation of SPDC photons along the normal direction.
Additional increase of the efficiency is expected by measuring the SPDC in transmission (as demonstrated by a 5-fold enhancement in the SHG efficiency~\cite{Camacho-Morales:2016-7191:NANL}), employing Fano-resonant nanostructures~\cite{Yang:2015-7388:NANL, Shorokhov:2016-4857:NANL,Vabishchevich:2018-1685:ACSP}, 
as well as by extending to an array of multiple nanoantennas, in the regime of spatial multiplexing with non-trivial spatial correlation  ~\cite{Wang1104,Stav1101}. 



In summary, we have experimentally demonstrated the generation of bi-photon states in an AlGaAs nanoantenna via SPDC. We have firstly inferred the polarization correlation of the generated photon pairs via the quantum-classical correspondence with the reversed SFG process. SFG polarization maps and radiation diagrams not only allowed us to determine the efficiency of nonlinear interactions in the resonator, but also reveal the directionality of the nonlinear emission. Secondly, we have demonstrated experimentally, for the first time to our knowledge, a spontaneous photon-pair source at the nanoscale with a bi-photon rate of up to 35~Hz. This is significantly higher than conventional SPDC photon sources, when normalised to the energy stored by the nanoantenna. We also predict further potential to increase the rates in combination with shaping the radiation pattern and frequency correlations in nanocylinder arrays.

The demonstrated nanoscale two-photon source lends itself to flexible quantum state engineering by possible tailoring of the spectral and radiation profile of the nanoantenna. We can also envisage the possibility of arranging multiple nanoantennas in a designer fashion to form a metasurface for generation of complex spatially entangled states. Our nanoantenna concept is not just limited to the SPDC photon-pair generation but is also applicable to other multi-photon quantum sources. It can thus open new opportunities for quantum applications such as free-space quantum communications.

\noindent \textbf{Supplementary Material} accompanies this paper.

\section*{Materials and Methods}

\subsection*{Sample Fabrication.}
We fabricate crystalline AlGaAs monolithic nanoantenna, since this material platform provides strong second order nonlinear susceptibility. 
The fabrication steps follow the procedure developed in Ref.~\cite{Gili:2016-15965:OE}. The AlGaAs layers are grown by molecular-beam-epitaxy on [100] non-intentionally doped GaAs wafer. 
A 400~nm layer of Al\textsubscript{0.18}Ga\textsubscript{0.82}As sits on top of a 1-$\mu$m-thick Al\textsubscript{0.98}Ga\textsubscript{0.02}As substrate sandwiched between two transition regions with varying aluminum molar fraction, in order to improve the eventual optical quality of the interface between AlOx and the adjacent crystalline layers. Patterned circles with radii of 215~nm, and equally spaced by $10~\mu$m, were produced with a scanning electron microscope lithography system. It followed a dry etching of the sample with non-selective ICP-RIE with SiCl\textsubscript{4}:Ar chemical treatment. The etching depth of 400~nm, controlled by laser interferometer, defined the nanocylinders and revealed the AlAs layer. The etched sample was then oxidized at 390$^{\circ}$C for 30 minutes in an oven equipped with in situ optical monitoring, under a precisely controlled water vapor flow with N\textsubscript{2}:H\textsubscript{2} gas carrier. After oxidation, each Al\textsubscript{0.18}Ga\textsubscript{0.82}As nanocylinder lies upon a uniform AlOx substrate, whose low refractive index enables sub-wavelength optical confinement in the nanocavity by total internal reflection. An SEM image of the AlGaAs nanocylinders is presented in Fig.~\ref{fig_concept}b. 

\subsection*{Numerical simulations.}
The linear and nonlinear response of the AlGaAs nanoantenna is modelled numerically using FEM solver in COMSOL Multiphysics, in the frequency domain. The material dispersion of AlGaAs is taken from Ref.~\cite{Gehrsitz:2000-7825:JAP}. 
The Q factor of the resonances are estimated by calculating the complex frequency of resonances around the fundamental and harmonics wavelengths using FEM solver through eigenfrequency analysis in COMSOL Multiphysics.
The second-order susceptibility tensor of the [100] grown AlGaAs, possessing a zinc blend crystalline structure, is anisotropic and contains only off-diagonal elements $\chi_{ijk}^{(2)}$ with $i\neq j\neq k$. We predict the SPDC output and correlations based on the quantum-classical analogy between the SPDC and the SFG processes~\cite{Poddubny:2016-123901:PRL}. For the SFG process, assuming an undepleted pump approximation, we follow two steps. Firstly, we simulate the linear scattering at the fundamental wavelengths $\lambda_{s}$ and $\lambda_{i}$. The bulk nonlinear polarization induced inside the particle is then employed as a source for the electromagnetic simulation to obtain the generated SF field. Based on the calculated SF field and the field at the signal and idler wavelengths $\lambda_{s}$ and $\lambda_{i}$, we further obtain the SPDC output correlations based on the quantum-classical analogy, as shown in Eq.~(\ref{eq:1}).

\subsection*{SFG measurements.}
Pulsed signal and idler beams are derived from a broadband femtosecond laser with a repetition rate of 80~MHz (Toptica, FibrePro). The two pulses are generated by spectrally slicing the 100~fs long pulses (bandwidth of 80 nm) into two paths at central wavelengths of 1520 and 1560~nm. After appropriate polarization conversion via the use of half and quarter wave plates, the two pulses are recombined with a 50:50 beam splitter and focussed onto the nanoantenna at normal incidence by a 0.7~NA objective (as shown in Fig.~S1 of the Supplementary Information). The reflected SFG radiation was collected in reflection through the same objective, separated from the signal and idler by a dichroic mirror. A short-pass filter at 800~nm is subsequently used for removal of the photoluminescence from the substrate, while a long-pass filter at 600~nm is used to remove the third harmonic emission component from the nanocylinder. The SFG emission was then acquired with a spectrometer or with a cooled camera in the real space. An additional confocal lens focusing at the objective back focal plane is used for imaging the emission pattern in the Fourier space.

\subsection*{SPDC measurements.}
The statistics of photons generated in SPDC can be characterised by measuring the second-order correlation $g^{(2)}=R_c / (R_1 R_2 \tau_c)$ using a beam splitter and two single-photon detectors at both outputs of this beam splitter ~\cite{Guo:2017-e16249:LSA}. Here $R_c$ is the rate of coincidences between the detectors, corresponding to SPDC, while $R_{acc}=R_1R_2\tau_c$ is the accidental coincidence rate that includes the count rates on each detector $R_1$ and $R_2$. The coincidence time window is $\tau_c$. We expect maximum $g^{(2)}$ at the zero time delay for the simultaneous arrival of the two photons~\cite{Klyshko:2011:QuantumElectronics}. 

In our experiments, we used a CW pump beam at a central wavelength of 785~nm and a linewidth of $<10$~MHz, horizontally polarized, to pump the nanoantenna at normal incidence via a 0.7~NA objective (as shown in Fig.~S4 of the Supplementary Information). 
The choice of the CW laser is justified by the fact that the SPDC photon-pair generation rate scales linearly with the average pump power. Furthermore, the CW operation allows us to eliminate the time-correlated noise if a pulsed pump was used, since a CW source has a flat temporal profile and a coherence time $>100$~ns, which is
larger than the measured coincidence time range of $\simeq 40$~ns. 

The reflected SPDC signal and idler photon pairs were collected in reflection through the same objective, separated by the pump with a dichroic mirror, and further filtered in free-space from the residual pump with the use of three long-pass filters at 1100~nm. The photon pairs were then separated by a 50:50 beam splitter into two paths and coincidences were measured with two gated InGaAs avalanche photo-diodes (IDQ230) and a time-tagging module (ID801-TDC). The detectors are coupled with multi-mode fibres and operate at $-90^{\circ}$C with an efficiency of 10\% and dark counts of 5~Hz. The counting scheme consisted of a coincidence window of 300 bins with a bin width of $\tau_c=162$~ps. 


\bibliography{sciadvbib}
\bibliographystyle{ScienceAdvances}

\noindent \textbf{Acknowledgements:} 
The authors acknowledge support from the Erasmus Mundus NANOPHI project (contract number 2013 5659/002-001) and the Australian Research Council (ARC) (DP150103733, DP160100619 and DE180100070). We acknowledge the use of the Australian National Fabrication Facility (ANFF) at the ACT Node. We thank Kai Wang, Matthew Parry, Frank Setzpfandt, Juergen Sautter, and Yuri Kivshar for the useful discussions. D.A.S. acknowledges partial support by the Russian President Grant No. MD-5791.2018. A.V.Z. acknowledges support from the ERC iCOMM project (789340), the Royal Society and the Wolfson Foundation. G.Z. acknowledges the financial support from NSFC (No. 11774182 and 91750204 ).

\end{document}